\documentclass[aps,prb,superscriptaddress,twocolumn,floatfix]{revtex4-2}

\usepackage{graphicx}
\usepackage[utf8]{inputenc}
\usepackage{color}
\usepackage{amsmath,amsfonts,amssymb}
\usepackage{lpic}
\usepackage{hyperref}
\usepackage[table]{xcolor}
\usepackage{soul}
\usepackage[normalem]{ulem}
\usepackage{pdfpages}
\usepackage{etoolbox} 

\makeatletter
\patchcmd{\@outputpage@head}{\@ifx{\LS@rot\@undefined}{}{\LS@rot}}{}{}{}
\makeatother

\definecolor{gray}{rgb}{.6,.6,.6}

\definecolor{auburn}{rgb}{0.43, 0.21, 0.1}

\definecolor{light-gray}{gray}{0.90}
\newcommand{\cellb}{\cellcolor{light-gray}}
\definecolor{darker-gray}{gray}{0.80}

\definecolor{dark-gray}{gray}{0.70}
\newcommand{\celld}{\cellcolor{dark-gray}}

\newcommand{\Eqref}[1]{Eq.~(\ref{#1})}
\newcommand{\Figref}[1]{Fig.~\ref{#1}}
\newcommand{\Secref}[1]{Sec.~\ref{#1}}

\newcommand{\dipc}{Donostia International Physics Center (DIPC), Paseo Manuel de Lardizabal 4, E-20018, Donostia-San Sebasti\'an, Spain}
\newcommand{\iker}{IKERBASQUE, Basque Foundation for Science, E-48013, Bilbao, Spain}
\newcommand{\cfm}{Centro de F\'{\i}sica de Materiales (CFM) CSIC-UPV/EHU, Paseo Manuel de Lardizabal 5, E-20018, Donostia-San~Sebasti\'an, Spain}

\newcommand{\ie}{\emph{i.e.}}
\newcommand{\eg}{\emph{e.g.}}

\newcommand{\SIangles}{S1}
\newcommand{\SIbondcurrents}{S2}
\newcommand{\SIsymmetries}{S3}
\newcommand{\SITAA}{S4}
\newcommand{\SITmatrices}{S5-S16}
\newcommand{\SIbandstructures}{S17}
\newcommand{\SIRTs}{S18}
\newcommand{\SIpertAA}{S19-S23}

\begin{document}

\title{Crossed graphene nanoribbons as beam splitters and mirrors \mbox{for electron quantum optics}}

\author{Sofia Sanz}
\email{sofia.sanz@dipc.org}
\affiliation{\dipc}

\author{Pedro Brandimarte}
\affiliation{\dipc}

\author{G\'eza Giedke}
\affiliation{\dipc}
\affiliation{\iker}

\author{Daniel S\'anchez-Portal}
\affiliation{\cfm}

\author{Thomas~Frederiksen}
\email{thomas\_frederiksen@ehu.eus}
\affiliation{\dipc}
\affiliation{\iker}

\date{\today}

\begin{abstract}
We analyze theoretically 4-terminal electronic devices composed of two crossed graphene nanoribbons (GNRs) and show that they can function as beam splitters or mirrors.
These features are identified for electrons in the low-energy region where a single valence or conduction band is present.
Our modeling is based on $p_z$ orbital tight-binding with Slater--Koster type matrix elements fitted to accurately reproduce the low-energy bands from density functional theory calculations.
We analyze systematically all devices that can be constructed with either zigzag or armchair GNRs in AA and AB stackings.
From Green's function theory the elastic electron transport properties are quantified as a function of the ribbon width.
We find that devices composed of relatively narrow zigzag GNRs and AA-stacked armchair GNRs are the most interesting candidates to realize electron beam splitters with a close to 50:50 ratio in the two outgoing terminals.
Structures with wider ribbons instead provide electron mirrors, where the electron wave is mostly transferred into the outgoing terminal of the other ribbon, or electron filters where the scattering depends sensitively on the wavelength of the propagating electron.
We also test the robustness of these transport properties against 
variations in intersection angle, stacking pattern, lattice deformation (uniaxial strain), inter-GNR separation, and electrostatic potential differences between the layers.
These generic features show that GNRs are interesting basic components to construct electronic quantum optical setups.
\end{abstract}

\maketitle

\section{Introduction}
\label{sec:introduction}

The similarities between the wave nature of electrons propagating coherently in ballistic conductors with photon propagation in optical wave guides has spawned the field of electron quantum optics \cite{EQO2017,Baeuerle18}.
In this way several electronic analogues of optical setups---such as the Mach--Zehnder \cite{Ji2003, Roulleau2007} and Fabry--P\'erot \cite{Zhang2009, McClure2009, Carbonell-Sanroma2017} interferometers, as well as the Hanbury Brown--Twiss \cite{Henny1999, Oliver1999, Samuelsson2004, Neder2007} geometry to study the Fermion antibunching and the two-particle Aharonov--Bohm \cite{Splettstoesser2010} effects---have been implemented.
Fundamental components for these setups include mirrors (M), beam splitters (BS, \ie, partially transparent mirrors), and wavelength filters.
Such control elements for electron beams are important in the fields of quantum information and solid-state quantum computation:
By sending a single electron through a BS one can generate a mode-entangled state that can be used to violate a Bell inequality \cite{Dasenbrook2016} or for quantum teleportation \cite{Bennett1993, Debarba2020}.
A BS is the central building block of the Hong--Ou--Mandel setup to test the indistinguishability \cite{Bocquillon2013} or the entanglement \cite{Giovannetti2006} of electrons entering in the two input ports.
With two BSs and two oriented Ms the Mach--Zehnder interferometer can be fully implemented, which has been demonstrated to work as a quantum logic processor \cite{Sarkar2006}.

A platform with remarkable prospects for electron quantum optics are graphene-based systems, in which several pioneering experiments on electron beam splitters and related devices have been performed \cite{RM+Schoenenberger15,Chen2016}.
More recently graphene nanoribbons (GNRs) \cite{Nakada1996, Fujita1996} have emerged as attractive candidates for the construction of molecular-scale electronic devices \cite{Schwierz2010} because they inherit some of the exceptional properties from graphene while having tunable electronic properties, such as the opening of a band gap depending on their width and edge topology \cite{Son2006a, Son2006b, Han2007, Yang2007, Wassmann2008}.
The electron coherence length in GNRs can be long, with values of the order $\sim 100$ nm being reported \cite{Minke2012,Baringhaus2014,Aprojanz2018}.
Furthermore, ballistic transport can be rather insensitive to edge defects because of the presence of localized edge states (\eg, in zigzag GNRs) and the dominating Dirac-like physics, that make the current flow maximally through the center of the ribbon \cite{Zarbo2007}.
With the advent of bottom-up fabrication techniques, long defect-free samples can be chemically synthesized with both armchair (AGNR) \cite{Cai2010} and zigzag (ZGNR) \cite{Ruffieux2016} edge topologies via on-surface synthesis.
Manipulation of GNRs with scanning tunneling probes has been also demonstrated \cite{Koch2012, Kawai2016}, opening the possibility to build two-dimensional multi-terminal graphene-based electronic circuits \cite{Areshkin2007, Jayasekera2007, Jiao2010, Cook2011, Botello-Mendez2011}.

Recently, it has been shown theoretically that two crossed GNRs with a relative angle of $60^\circ$ can behave as a BS for valence- and conduction-band electrons \cite{Lima2016, Brandimarte2017}, since such four-terminal devices were found to divide the electron beam into two out of the four arms with an equal transmission probability of $50\%$.
In this paper we analyze this possibility more generally and show that all the mentioned beam-control elements (BS, M, filters) can be realized with a suitable choice of two crossed GNRs.
More specifically, we compute the electron transport properties of these devices in terms of the edge topology and width of the GNRs, and the precise alignment and stacking of the ribbons.

The problem is theoretically approached by means of tight-binding (TB) modeling, which is known to reproduce graphite-like systems with sufficient accuracy \cite{Wallace1947, Dresselhaus2002, Reich2002, McCann2006, Malard2007}, while allowing to explore a large number of systems of considerable sizes in a fast and transparent way.
For instance, the geometry of a crossing between two 50-atom wide GNRs readily comprises around $10\,000$ atoms.
The main complexity of the modeling lies in the description of the interlayer couplings, for which we use a Slater--Koster parametrization \cite{Slater1954} that has proven successful for describing the band structure and velocity renormalization of Dirac electrons in twisted bilayer graphene \cite{LoPeCa.07.GrapheneBilayerwith, TramblydeLaissardiere2010}.
The employed technique can describe arbitrary device geometries and therefore allows us to also study of the robustness of the predicted transport properties against variations in intersection angle, stacking pattern, lattice deformation (uniaxial strain), inter-GNR separation, and electrostatic potential differences between the layers.
With this, we give a complete analysis of the transport properties of crossed GNRs, highlight their tunability, and provide quantitative data that can serve as a guide for design optimization.

This paper is organized as follows: in \Secref{sec:methodology} we introduce the general TB Hamiltonian used to describe the kinetics of the electrons travelling through the different devices as well as the scattering formalism used to compute transmission and reflection probabilities of incoming electron waves from the different leads.
In \Secref{sec:results} we present a complete analysis of the transport properties based on the key combinations of stacking pattern, edge topology and width of the GNRs.
Finally, the conclusions and remarks are provided in \Secref{sec:conclusion}.

\section{Methodology}
\label{sec:methodology}

\begin{figure}[t]
    \includegraphics[width=\columnwidth]{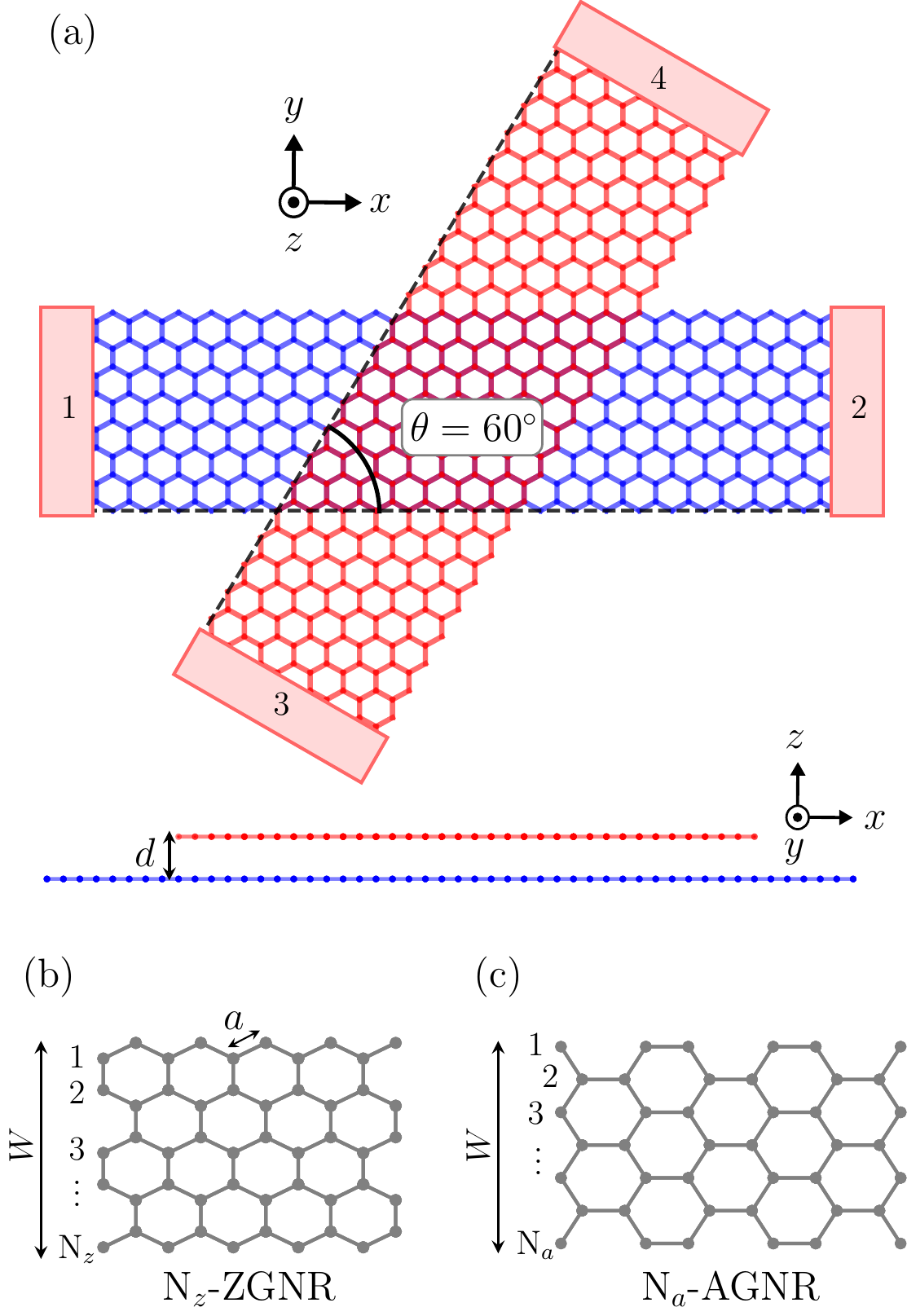}
    \caption{Illustration of the general setup.
    (a) A four-terminal device is formed by two crossed 8-ZGNRs with a relative angle $\theta=60^\circ$.
	The bottom (top) ribbon is drawn in blue (red) with carbon atoms at each vertex.
	The four semi-infinite leads, numbered 1 to 4, are attached in the contact regions represented with red rectangles. The ribbons lie out-of-plane separated by a distance $d$ along the $z$-axis (see side view).
	Definition of the width $W$ of (b) ZGNRs and (c) AGNR in terms of the number of carbon atoms $N$ across the ribbon.
	The interatomic distance is denoted by $a$.
    }
	\label{fig:definition-terminals}
\end{figure}

The general setup of this study, illustrated in \Figref{fig:definition-terminals}a, is comprised of two infinite GNRs crossed with a relative angle $\theta=60^{\circ}$ (see Sec. III. A for a discussion on this choice of angle).
The scattering region (intersection) breaks the translational invariance of the infinite ribbons, for which we will use the Green's function formalism to solve the Schr\"odinger equation with open boundary conditions.

The system is divided into the device (scattering) region that contains the intersection between the two ribbons, and the four semi-infinite GNRs (periodic electrodes), represented as red rectangles in \Figref{fig:definition-terminals}a.
The total Hamiltonian is correspondingly split into the different parts
\begin{equation}
H_{T} =  H_{d} + \sum_{\alpha} (H_{\alpha} + H_{\alpha d}),
\end{equation}
where $H_{d}$ is the device Hamiltonian, $H_{\alpha}$ the $\alpha$-electrode Hamiltonian, and $H_{\alpha d}$ the coupling between these two subsystems.

\subsection{Tight-binding Hamiltonian}
	
The use of a local basis in combination with Green's function techniques provides an efficient way for obtaining the transport properties in terms of microscopic parameters.
We write the single-particle TB Hamiltonian in an orthogonal basis as
\begin{equation}\label{eq:TB-Hamiltonian}
	H = \sum_{i}\epsilon_{i}c^{\dagger}_{i}c_{i} + \sum_{ij}t_{ij}\left(c^{\dagger}_{i}c_{j} + h.c. \right),
\end{equation}
where $c^{\dagger}_{i}$ ($c_{i}$) creates (annihilates) an electron on site $i$ with energy $\epsilon_i$.
We further define the Fermi level as $E_F=\epsilon_i$, corresponding to half-filled carbon $p_z$ orbitals.
The matrix element $t_{ij}$ between orbitals $i$ and $j$ is described by Slater--Koster type two-centre $\pi$ and $\sigma$ bond integrals between two $p_{z}$ atomic orbitals \cite{Slater1954}
\begin{equation}\label{eq:Slater-Koster}
	t_{ij} = V_{pp\pi}(1-l^{2}) + V_{pp\sigma}l^{2},
\end{equation}
where $l$ is the cosine of the angle formed between the distance vector $\hat{r}_{ij}$ for the $ij$ atom pair and the unit vector that defines the $z$-direction (cf.~\Figref{fig:definition-terminals}a), \ie, $l = {\hat{r}_{ij}\cdot\hat{e}_{z}}/{|r_{ij}|}$.
The two-centre integrals are expressed as
\begin{equation}\label{eq:Vpi}
	V_{pp\pi} = -t_{\parallel}e^{q_{\pi}\left(1-\frac{|r_{ij}|}{a}\right)},
\end{equation}
\begin{equation}\label{eq:Vpsig}
	V_{pp\sigma} = -t_{\perp}e^{q_{\sigma}\left(1-\frac{|r_{ij}|}{d}\right)},
\end{equation}	
where $t_{\parallel}$ ($t_{\perp}$) is the \emph{intra}-GNR (\emph{inter}-GNR) hopping parameter between atoms separated by the interatomic (interlayer) distance fixed to $a=1.42$ {\AA} ($d=3.34$ {\AA}) in our model, see \Figref{fig:definition-terminals}.
The decay rates of the bond integrals with the atomic separation, $q_{\sigma}$ and $q_{\pi}$, are isotropic and therefore related by ${q_{\sigma}}/{d}={q_{\pi}}/{a}$.
This model, characterized by $t_{\parallel}, t_{\perp}$ and the decay rate (which can be determined by fixing the second nearest neighbors coupling), successfully describes $\pi$ electrons in twisted bilayer graphene \cite{TramblydeLaissardiere2010}.
However, it does not capture many-body effects like, \eg, the difference in nearest-neighbor hopping parameter for different lattice sites as in the Slonczewski--Weiss--McClure (SWM) model for graphite \cite{Dresselhaus2002, Malard2009, CastroNeto2009, Guinea2019}.

In this work we use $t_{\parallel}=2.682$ eV and $t_{\perp}=0.371$ eV.
For the third model parameter we refer to the in-plane next-nearest neighbor matrix element $t^{\prime}=0.0027$ eV.
These parameters were obtained by fitting to the low-energy band structure of AB-stacked bilayer graphene simulated with \textsc{Siesta} \cite{Soler2002} as explained in Appendix \ref{sec:supp-DFT-fitting}.
The satisfactory agreement between TB and DFT (\Figref{fig:supp-DFT-vs-TB}) further justifies that, at least for our purposes, many-body effects like in the SWM model can be neglected.

\subsection{Transport calculations}

In order to perform transport calculations we use the nonequilibrium Green's function (NEGF) method \cite{Kadanoff1962, Keldysh1965, HaJa.07.QuantumKineticsin}.  In particular, to obtain the transmission probabilities ($T_{\alpha\beta}$) between the different pairs of electrodes ($\alpha\neq \beta$), we use the Landauer-B\"uttiker formula \cite{Buettiker1985}, 
\begin{equation}\label{eq:T-probability}
    T_{\alpha\beta} = \mathrm{Tr}\left[\Gamma_{\alpha}G\Gamma_{\beta}G^{\dagger}\right], \qquad \alpha \neq  \beta
\end{equation}
where $\Gamma_{\alpha} = i (\Sigma_{\alpha} - \Sigma^{\dagger}_{\alpha})$ is the broadening matrix, related to the nonhermitian part of the retarded electrode self-energy $\Sigma_\alpha$, due to the coupling of the $\alpha$th semi-infinite lead to the scattering center and $\alpha,\beta=1,\dots,4$, cf.~\Figref{fig:definition-terminals}.
Further,
\begin{equation}\label{eq:dev-G}
   G_{d} = \left(\mathbb{I}E-H_{d}-\sum_{\alpha}\Sigma_{\alpha}\right)^{-1}
\end{equation}
is the retarded Green's function of the device region and $\mathbb{I}$ the identity matrix (orthogonal basis).
The dependency on the electron energy $E$ of these key quantities is implicit.

The reflection probability ($T_{\alpha\alpha}=R_{\alpha}$) can be conveniently written as a difference between the bulk electrode transmission $M_{\alpha}$ (\ie, the number of open channels/modes in electrode $\alpha$ at a given energy) and the scattered part into the other electrodes ($\sum_{\beta}T_{\alpha\beta}$) as
\begin{equation}\label{eq:R-probability}
   R_{\alpha} = M_{\alpha} - \sum_{\beta\neq\alpha}T_{\alpha\beta}.
\end{equation}
From \Eqref{eq:dev-G} we can also obtain the spectral function $A_\alpha$ for states coupled to electrode $\alpha$
\begin{equation}\label{eq:ADOS}
	A_{\alpha} = G\Gamma_{\alpha}G^{\dagger}.
\end{equation}	  
The diagonal elements $A_{\alpha}(i,i)/2\pi$ correspond to the local density of states (DOS) at sites $i$ of the scattering states originating from electrode $\alpha$.

Computationally, we constructed the Hamiltonian matrix with the \textsc{sisl} package \cite{zerothi_sisl, Papior2017} and computed transmission probabilities and spectral DOS with \textsc{TBTrans} \cite{Papior2017}.

\section{Results}
\label{sec:results}

In this section we present results for the electron transport properties through an extensive set of four-terminal devices formed of two crossed ribbons.
We analyze the role of the precise stacking and alignment of the crossing area for both ZGNR- and AGNR-based devices in all their possible configurations.
	
\begin{figure}
	\includegraphics[width=\columnwidth]{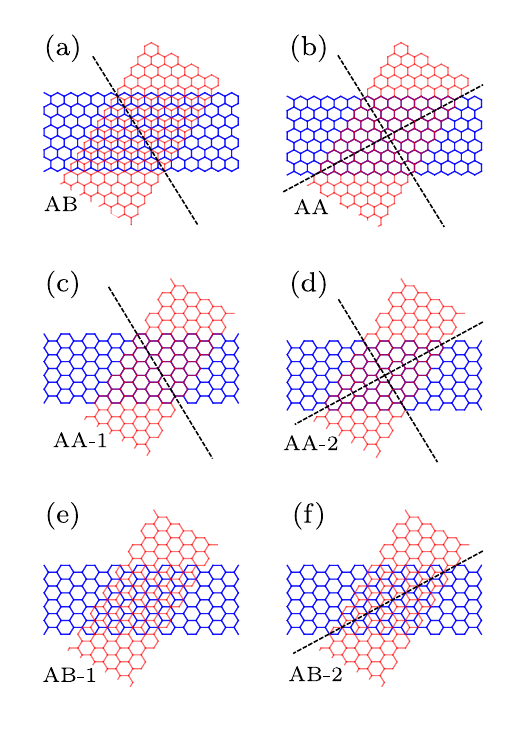}
	\caption{Geometries of the different stackings that can be constructed from the crossing of two GNRs with a relative angle of $60^\circ$.
	The bottom (top) ribbon is drawn in blue (red) with carbon atoms at each vertex.
	For ZGNR-based devices there exist only one AA- and one AB-stacked configuration, labeled (a) \texttt{AB} and (b) \texttt{AA} (exemplified here by 8-ZGNR).
	For AGNR-based devices there exist two AA- and two AB-stacked configurations, labeled (c) \texttt{AA-1}, (d) \texttt{AA-2}, (e) \texttt{AB-1}, and (f) \texttt{AB-2} (exemplified here by 11-AGNR).
	The dashed lines show the symmetry (reflection) planes that preserve the Hamiltonian of each crossing when such operation is applied to them.
	}
	\label{fig:devices}
\end{figure}

\subsection{Possible device configurations}
\label{sec:devices}

\begin{figure}
	\includegraphics[width=\columnwidth]{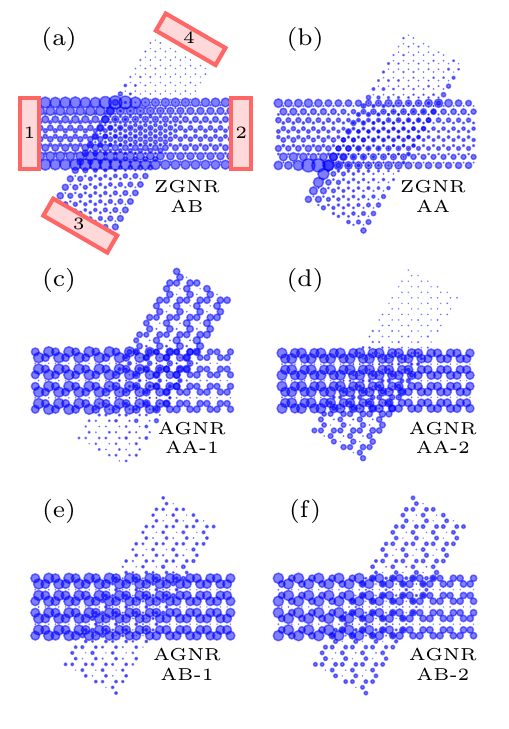}
	\caption{Spectral DOS of scattering electrons incoming from electrode $\alpha=1$ obtained from \Eqref{eq:ADOS}, for the specific geometries defined in \Figref{fig:devices}:
	(a) 8-ZGNR \texttt{AB}, (b) 8-ZGNR \texttt{AA}, (c) 11-AGNR \texttt{AA-1}, (d) 11-AGNR \texttt{AA-2}, (e) 11-AGNR \texttt{AB-1}, and (f) 11-AGNR \texttt{AB-2}.
	The spectral DOS were calculated at $E=200$ meV for ZGNRs and at $E=0$ meV for AGNRs.
	}
	\label{fig:resolved-ADOS}
\end{figure}

The symmetry of the honeycomb lattice yields a perfect matching between the bottom and top GNR lattices for $\theta=60^\circ$.
In this situation is expected that the maximized interlayer coupling enhances the transfer of electrons between the ribbons, as shown in \cite{Botello-Mendez2011, Lima2016, Brandimarte2017}.
In Fig. \SIangles in the Supplemental Material (SM) \cite{SM} we performed transport caclulations for crossed 8-ZGNRs both in the AA- and AB-stackings as a function of the crossing angle between the GNRs, where such behavior is observed for angles approaching $60^\circ$.
We therefore focus the discussion on devices formed by crossed GNRs with an intersecting angle of $\theta=60^\circ$.
However, the \textit{inter}-GNR transmission is also enhanced for angles within $[50^\circ,70^\circ]$, which highlights the tunability of our devices.
 Note that experiments on twisted bilayer graphene report that the rotation angle between the layers can be precisely controlled down to fractions of a degree (0.01$^\circ$) \cite{Li2010, Cao2018a, Cao2018b}.

For a systematic analysis we begin by considering all the possible devices that can be built with two crossed AA- or AB-stacked GNRs with a relative angle of $60^\circ$.
These are sketched \Figref{fig:devices}.
In the case of crossed ZGNRs there exist two configurations, the AB-stacking [labeled \texttt{AB}, \Figref{fig:devices}(a)] and the AA-stacking [labeled \texttt{AA}, \Figref{fig:devices}(b)].
These two geometries have different symmetries, indicated by the reflection planes (dashed lines) in \Figref{fig:devices}.
While \texttt{AB} has only one reflection symmetry, \texttt{AA} has two.
Here, and in the following, we refer only to symmetries in the $xy$-plane.
The additional operation of reflection in the $z$-direction to interchange top and bottom ribbon is physically not important and therefore implicit.

In the case of AGNRs there are two different AA-stacked configurations [labeled \texttt{AA-1} and \texttt{AA-2}, \Figref{fig:devices}(c,d)], as well as two different AB-stacked configurations [labeled \texttt{AB-1} and \texttt{AB-2}, \Figref{fig:devices}(e,f)].
For instance, starting from \texttt{AA-1}, one can obtain \texttt{AA-2} by translating the upper (red) ribbon by $-\sqrt{3}a\hat{y}$ with respect to the lower one.
Similarly, \texttt{AB-1} (\texttt{AB-2}) can be obtained from \texttt{AA-1} by translating the upper (red) ribbon by $+a\hat{x}$ ($-a\hat{x}$) with respect to the lower one.
Again, these four generic configurations have different symmetries as indicated in \Figref{fig:devices}(c-f).

The reflection planes imply that there are operations which leave the scattering potential (created by the intersection of the two ribbons) unchanged.
This is, if we apply one or more reflections across the indicated axes, the Hamiltonian of the new device does not change.
Consequently, the Green's function and all the transport properties derived from it, will also remain unchanged under some particular electrode permutations.

Let us begin by discussing the properties of these six different configurations with particular examples constructed from 8-ZGNRs and 11-AGNRs.
In \Figref{fig:resolved-ADOS} we show the spectral DOS of scattering electrons that come in from electrode $\alpha=1$ as obtained from \Eqref{eq:ADOS} for each configuration at specific energies.
In this real-space representation it is easy to see where the scattered electron wave propagates after being injected into the device.
The large DOS that appears in the input electrode region does not correspond to the backscattered electrons, but rather to the DOS of the incoming electrons (as we will show later on).
This is also illustrated in Fig.~{\SIbondcurrents} \cite{SM}, where we complement the results shown in \Figref{fig:resolved-ADOS} by plotting the bond currents between nearest neighbor atoms, where the arrows indicate the direction of the flowing electrons.

For the ZGNR devices, \Figref{fig:resolved-ADOS}(a,b) and Fig.~{\SIbondcurrents}(a,b) show that an electron injected from $\alpha=1$ in both cases only escapes from the scattering center into electrodes $\beta=2,3$, \ie, terminals $1$ and $4$ are suppressed.
This lack of backscattering (and preferential scattering into only one of the two arms of the other ribbon) is a very general and robust feature for ZGNRs which holds for different widths, stackings, and energies and it is instrumental for the applications we have in mind.
An explanation, supported by continuum-model calculations \cite{Brey2006, Wakabayashi2007}, is the valley (chirality) preservation in low-energy bands of ZGNRs.
For the two AA-stacked AGNR devices \Figref{fig:resolved-ADOS}(c,d) and Fig.~{\SIbondcurrents}(c,d) show that the outgoing terminals $\beta=1$ and $\beta=3$ ($\beta=4$) for \texttt{AA-1} (\texttt{AA-2}) are suppressed.
These two cases are interesting since their relative displacement of only $\sqrt{3}a\hat{y}$ leads to very different electron transport: for \texttt{AA-1} the split electron turns by $60^\circ$, while for \texttt{AA-2} the bend is $120^\circ$.
Unlike for ZGNR devices, the suppression of two terminals is not general for all AGNR widths, and rather depends on the AGNR family, as shown in Figs.~\SITmatrices \cite{SM}.
In the case of the two AB-stacked ribbons, \Figref{fig:resolved-ADOS}(e,f) show that an electron wave in these devices is scattered qualitatively (yet not quantitatively) similarly and into all outgoing electrodes.

\subsection{Symmetry considerations}

Since we deal with 4-terminal devices, the matrix of transmission and reflection probabilities, \Eqref{eq:T-probability} and \Eqref{eq:R-probability}, has the general form
\begin{equation}\label{eq:Tpq-matrix}
   \textbf{T} = \begin{pmatrix}
    R_{1}        & T_{12}        & T_{13}        & T_{14} \\
   \celld T_{21} & \cellb R_{2}  & \cellb T_{23} & \cellb T_{24} \\
   \celld T_{31} & \celld T_{32} & \cellb R_{3}  & \cellb T_{34} \\   
   \celld T_{41} & \celld T_{42} & \celld T_{43} & \cellb R_{4} \\
  \end{pmatrix}.
\end{equation}
However, due to symmetries there are not 16 independent quantities in this matrix.
First of all, in absence of a magnetic field, time reversal symmetry forces $T_{\alpha\beta}=T_{\beta\alpha}$.
This reduces the matrix to 10 independent elements, \eg, those without the dark gray background ($\alpha>\beta$) in \Eqref{eq:Tpq-matrix}.
Secondly, the symmetries indicated in \Figref{fig:devices} reduce the number of independent elements of the matrix further.
The reflection plane $y=\sin(-60^\circ) x$ maps the electrode labels (1, 2, 3, 4) $\leftrightarrow$ (4, 3, 2, 1) with unchanged transmissions, 
\eg, which allows to consider $R_3$, $R_4$, $T_{24}$ and $T_{34}$ as dependent variables [4 of the light gray elements in \Eqref{eq:Tpq-matrix}].
Similarly, the reflection plane $y=\sin(30^\circ)x$ implies (1, 2, 3, 4) $\leftrightarrow$ (3, 4, 1, 2)
and $R_3$, $R_4$, $T_{23}$, and $T_{34}$ as possible dependent variables (4 of the light gray elements).
The combination of both reflection planes further implies (1, 2, 3, 4) $\leftrightarrow$ (2, 1, 4, 3)
and $R_2$ and $T_{23}$ as further dependent variables (\ie, all gray elements in this case).
In summary, depending on the number of symmetries, the transmission probabilities of any given device will be fully characterized by either 4, 6 or 10 independent matrix elements.

\begin{figure}
	\includegraphics[width=\columnwidth]{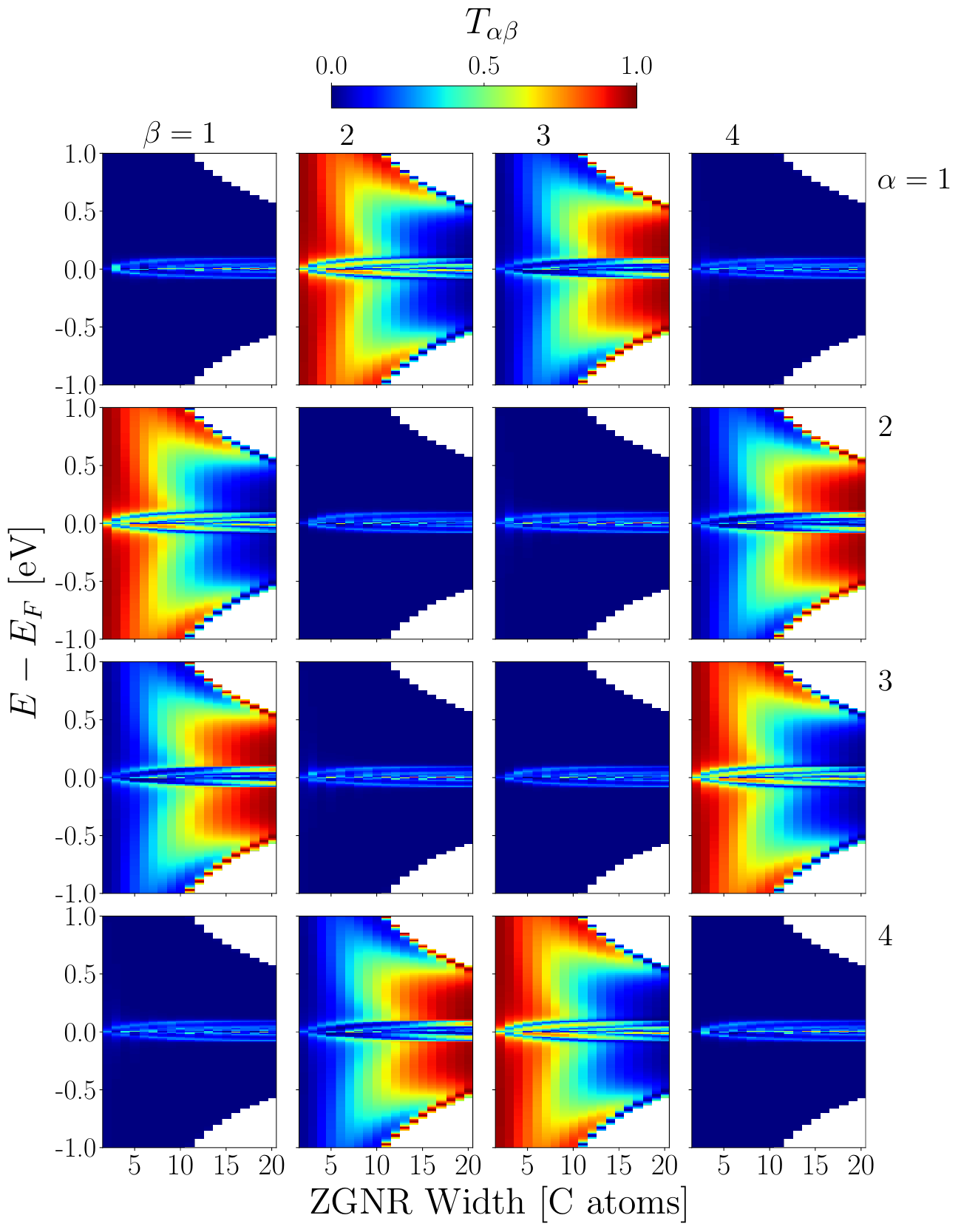}
	\caption{Full transmission probability matrix $T_{\alpha\beta}$ between all the electrode pairs for ZGNRs crossed in the AB configuration as a function of the ribbon width $W$ and electron energy $E$.
	Only data for the first subband is shown (white regions correspond to multiple electronic bands in the ribbons).
	}
	\label{fig:Total-Tmatrix}
\end{figure}
	
Figure \ref{fig:Total-Tmatrix} shows the full, energy-resolved transmission matrix [\Eqref{eq:Tpq-matrix}] obtained numerically for devices formed of two crossed ZGNRs in the \texttt{AB} configuration for a range of different ribbon widths $W$.
As ZGNR \texttt{AB} displays only one reflection plane, the transmission probabilities for these systems are, in principle, characterized by 6 independent quantities.
However, qualitatively only 4 independent ones are readily identified in \Figref{fig:Total-Tmatrix}.
Only upon close inspection of the data all the expected differences emerge.
The reason for the seemingly higher symmetry (corresponding to two reflection planes) is the fact that the scattering potential created by the crossings between the GNRs, depends exponentially on the atomic distances between the GNRs and, therefore, is dominated by the closest atom pairs.
These atom pairs, shown in Fig.~{\SIsymmetries}(a) \cite{SM}, are in fact symmetric with respect to both reflection planes.

More generally, for all the configurations in \Figref{fig:devices} we find that the scattering potentials are dominated by terms with at least one reflection plane (Fig.~{\SIsymmetries}).
For all practical purposes, the effective symmetry appears higher and it suffices to describe the transmission probabilities with only 4 or 6 independent quantities.

In the following we will thus only consider it sufficient to discuss electrons incoming from terminal $\alpha=1$.
However, for completeness we show the full transmission matrices for all the considered systems in Figs.~{\SITAA}-S16 \cite{SM}.

\subsection{Beam splitters and mirrors}
Looking again at \Figref{fig:Total-Tmatrix} and focusing on the first row (electron beam injected from terminal $\alpha=1$), we observe distinct regimes where the devices would present particular electron quantum optical characteristics.
We are especially interested in geometries for which the transmission matrix allows to designate two input and two output electrodes in the sense that any electron sent in through one of the input ports is scattered predominantly into the two output ports with very little reflection or transmission into the other input.
For instance, the green areas in the plots show where the device behaves as a BS, since they show that the electron beam is scattered only into \emph{two} out of the four possible arms with a transmission probability that lies around $T_{12}\sim T_{13}\sim 0.5$.
One can also identify regimes in which the device can work as a M where $T_{13}\sim 1$. 
This situation corresponds to the red areas in \Figref{fig:Total-Tmatrix}, since the electron would enter from terminal $\alpha=1$ and turn $120^\circ$ to go out \emph{exclusively} into terminal $\beta=3$ with low reflection.
The energy-dependence of the transmission functions is very symmetric with respect to the Fermi level, reflecting the approximate particle-hole symmetry characteristic of a half-filled bipartite lattice.
Nevertheless, the presence of next-nearest neighbor couplings in our TB model in principle breaks this symmetry.

On one hand, we note that for energies close to the Fermi level ($|E-E_{F}| < 0.07$ eV) in \Figref{fig:Total-Tmatrix}, the electron is scattered into all the four output ports, which makes this small energy window not very interesting for electron quantum optical purposes.
These features probably arise due to the hybridization of states
from the flat bands of the individual ribbons in the overlapping area.
The band structures for both monolayer and bilayer ZGNRs are shown in Fig.~{\SIbandstructures} \cite{SM}.
On the other hand, we note here that outside the low-energy region (where there is more than one electronic band) we find for all systems that reflection and interband scattering play a larger role in the electron transport through these devices, as the number of open channels (modes) grows with energy.
In other words, it was not possible to identify conditions for realizing BS or M at energies with multiple subbands in the GNRs.
Therefore the following discussion is focused on the energy window corresponding to a single (conduction or valence) band, since the most interesting physics related to the electron quantum optical features were identified here. 

\subsection{Quality of the realized mirrors and beam splitters}

\begin{figure}
    \includegraphics[width=\columnwidth]{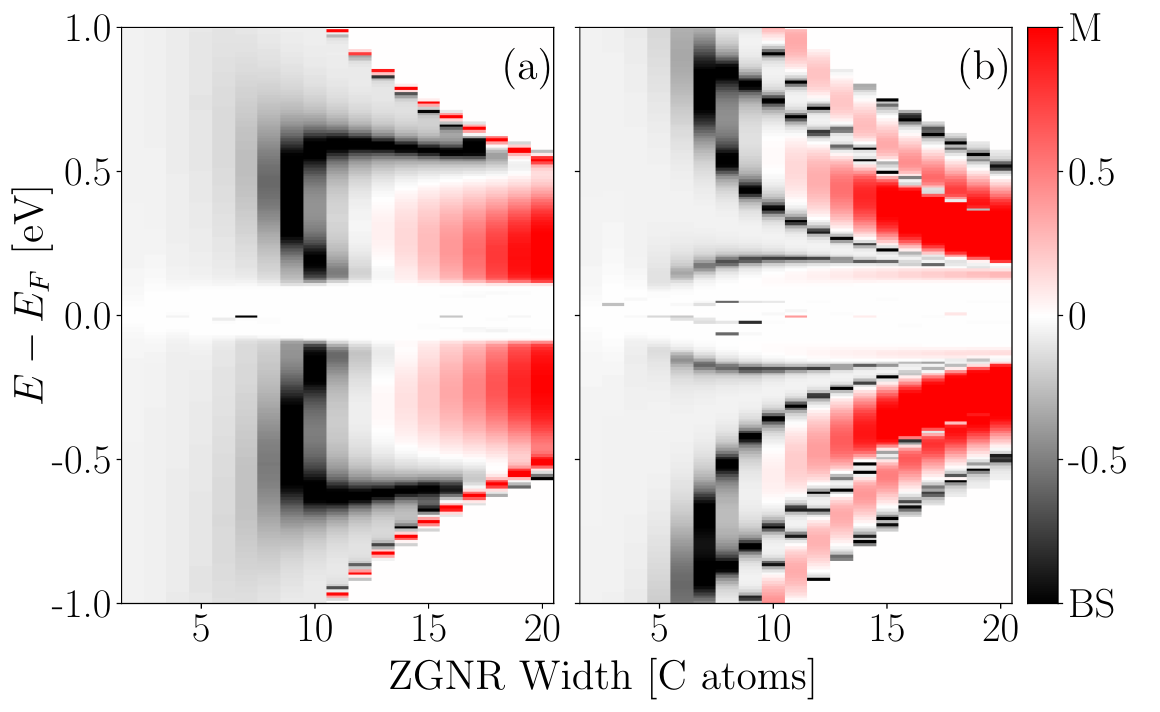}
	\caption{Figure of merit (FM) for systems composed of ZGNRs
	in (a) \texttt{AB} or (b) \texttt{AA} configurations.
	Black and red regions correspond to situations where a given device is suitable as a BS or M, respectively.
	White regions are unsuitable as BS or M because of large transmission into the other but the desired output ports.
    }
    \label{fig:ZGNR-T-enhanced}
\end{figure}

\begin{figure}[h!]
	\includegraphics[width=\columnwidth]{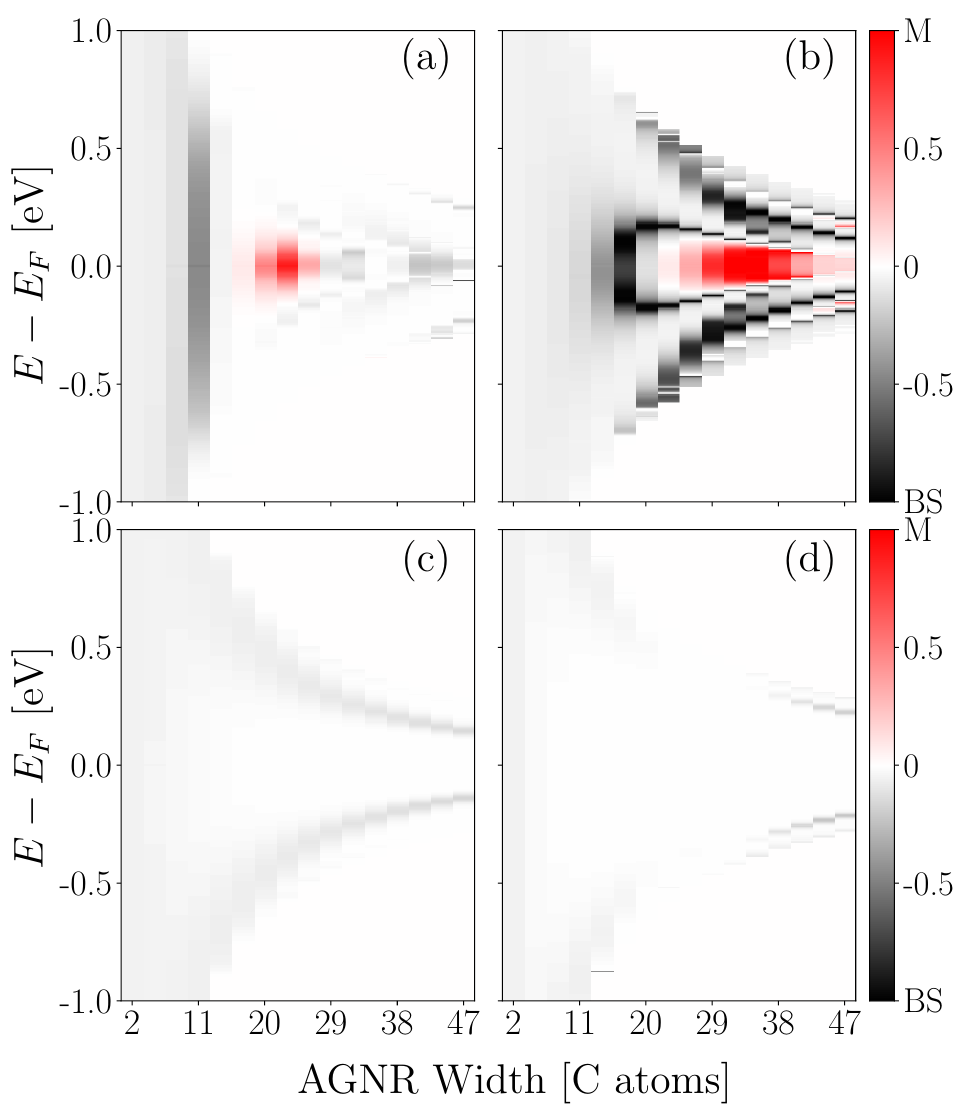}
	\caption{
	Figure of merit (FM) for systems composed of AGNRs of the $3p+2$ family
	with the (a) \texttt{AA-1}, (b) \texttt{AA-2}, (c) \texttt{AB-1}, and (d) \texttt{AB-2} configurations.
	Black and red regions correspond to situations where a given device is suitable as a BS or M, respectively.
	White regions are unsuitable as BS or M because of large transmission into the other but the desired output ports.
    }
    \label{fig:AGNR-T-enhanced}
\end{figure}

\begin{figure*}[t]
	\includegraphics[width=\textwidth]{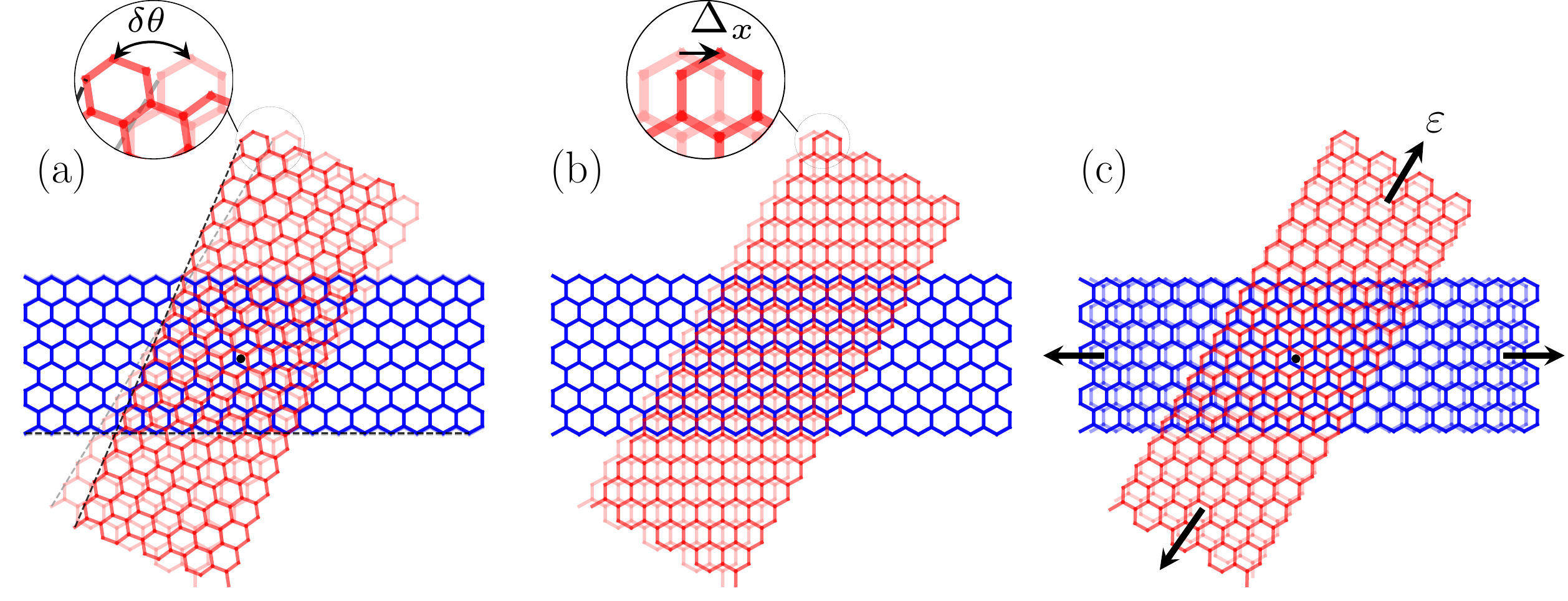}
	\caption{Sketch of the geometrical distortions applied to the AB-stacked 8-ZGNR device.
	(a) Rotation by some small angle $\delta\theta$ around the configuration with relative angle of $\theta=60^\circ$. The rotation is performed around the center of the scattering region, indicated by a black dot.
	(b) Translation of the on-top ribbon with respect to the lower one by an amount $\Delta_{x}$ along the $x$-axis.
	(c) Strain $\varepsilon$ is applied along the periodic direction of each ribbon while keeping the center of the scattering (black dot) region unchanged.
    }
    \label{fig:geometrical-distortions}
\end{figure*}

To obtain a qualitative picture across all the possible systems of the most suitable candidates for BSs or Ms, we construct in the following a figure of merit (FM).
On the one hand, we look for candidate systems where a significant part of the scattered electron wave can be transferred to the other ribbon, \ie, that $T_{13}$ or $T_{14}$ is large.
We encode this property in the quantity $\tau \equiv \max(T_{13}, T_{14})$.
On the other hand, for a suitable BS or M it is important that the reflection \emph{and} transmission to a third electrode should be small.
This property is encoded as a "loss" function $\lambda \equiv R_{1}+\min(T_{12}, T_{13}, T_{14})$.

Our FM is then defined as 
\begin{equation}\label{eq:FM-definition}
    \mathrm{FM} = e^{-20\lambda}\tanh\left[ \frac{1}{20}\left(\frac{1}{|\tau -1|} - \frac{1}{|\tau -1/2|} \right) \right].
\end{equation}
We use a linear color scale where BSs ($\mathrm{FM}=-1$) appear as black,  Ms ($\mathrm{FM}=1$) as red, and uninteresting systems ($\mathrm{FM}=0$) as white.
We set FM equal to zero whenever there is more then one band per GNR at the energy considered (as it happens, e.g., for large values of $|E-E_F|$).
In that case the sum of all transmission probabilities is equal to the number of bands and thus larger than 1. This case is not useful for the devices we have in mind, though a more careful analysis may show how to also use the systems in this energy range.
In other words, $\lambda$ determines the intensity of the plots while $\tau$ sets the color.
The FM is chosen to be highly selective: it decays to about 1/2 of the maximum value for losses (=transmission probability into the undesired output ports) of about 3\%.
Similarly, the FM of a loss-free, but unbalanced BS is reduced to $\mathrm{FM}=-1/2$ at an imbalance of about 57:43.
Figures \ref{fig:ZGNR-T-enhanced} and \ref{fig:AGNR-T-enhanced} show the FM for ZGNRs and AGNRs from the metallic $3p+2$ family, respectively, as a function of electron energy and ribbon width $W$.
Overall, these figures show that the most interesting systems are those composed by ZGNRs or AA-stacked AGNRs.
For both types of GNRs one can find devices that behave as BS or as M, respectively.
For instance, \Figref{fig:ZGNR-T-enhanced} reflects that the 8-ZGNR devices shown in \Figref{fig:devices}(a,b) are good candidates for BS, consistent with the qualitative picture of \Figref{fig:resolved-ADOS}(a,b) and \Figref{fig:Total-Tmatrix}.

For both \texttt{AA} and \texttt{AB} ZGNR devices the transmitted electron wave to the other ribbon is always bent 120$^\circ$ into electrode 3 (see also the full transmission matrices in \Figref{fig:ZGNR-T-enhanced} and Fig.~{\SITAA} \cite{SM}).
To obtain a M, where an electron incoming from electrode 1 is almost entirely transferred to electrode 3, one should choose wider ZGNRs.

For the AGNRs the situation is a little more complex.
As discussed in \Figref{fig:devices} it is possible to form four different stackings (\texttt{AA-1}, \texttt{AA-2}, \texttt{AB-1} and \texttt{AB-2}).
Further, the band gap of AGNRs is determined by the overall ribbon width $W$, which classifies them into three distinct families $3p$, $3p+1$, or $3p+2$ for integer $p$ \cite{Son2006a, Son2006b, Yang2007, Wassmann2008}.
This leaves us with 12 different situations, considered in terms of the full transmission matrices in Figs.~{\SITmatrices} \cite{SM}.
We find that the most interesting devices are those 
built with $(3p+2)$-AGNRs in the AA-stacked configurations.
However, compared with the ZGNRs, the parameter space for desirable devices is more restricted and the losses are generally larger.
Independent of width, the AB-stacked configurations lead to scattering of the electron wave into all terminals.

We also note here that the qualitative difference mentioned in \Secref{sec:devices} between the 60$^\circ$ turn of the transferred electron wave for \texttt{AA-1} configuration versus the $120^\circ$ turn for \texttt{AA-2} is a robust feature across the different families (Figs.~{\SITmatrices} \cite{SM}).
Additionally, we also find very thin white regions, that do not correspond to high losses but to $T_{12}\sim 1$, immersed in red -- \eg, seen for $W=10$-$15$ in \Figref{fig:ZGNR-T-enhanced}(b) and for $W>20$ in \Figref{fig:AGNR-T-enhanced}(b). 
This suggests that crossed GNRs can also work as energy filters.
These $T_{12}$ ($T_{13}$) peaks (dips), also plotted in Fig.~{\SIRTs} for clarification, become narrower as the width of the ribbons grows, which enhances the energy selection.

\subsection{Robustness of the discussed properties}
\label{sec:robustness}
So far we have discussed the different transport properties that can be found in the ideal case, that is commensurate GNRs (AA- or AB-stacking) with a relative angle of $\theta=60^{\circ}$.
However, precise control of the device geometry is likely a significant experimental challenge.
In this section we therefore proceed to test the robustness of the transport properties against some perturbations of this ideal situation.
More specifically, we explore now the exact roles of the intersection angle, deviations from the idealized stacking pattern, lattice deformations via uniaxial strain, variation of the inter-GNR separation, and electrostatic potential differences between the two ribbons.

Since we concluded above that ZGNR devices may be the best candidates for building electron quantum optical setups, we will focus the following discussion around them.
We take as the reference device the crossing of two AB-stacked 8-ZGNRs (\Figref{fig:devices}a) and compute the transmission probabilities from terminals $\alpha=1$ to $\beta=1,2,3,4$ for each of the above mentioned perturbations.
The AA-stacked 8-ZGNRs were found to display qualitatively similar trends
as can be seen from the Figs.~{\SIpertAA} \cite{SM}.
We will see that the low-loss property of these devices is thus preserved for the applied variations and in some cases the FM is even significantly enhanced, indicating that almost perfect BS or M could be obtained by tuning the above mentioned parameters.

\begin{figure}
	\includegraphics[width=\columnwidth]{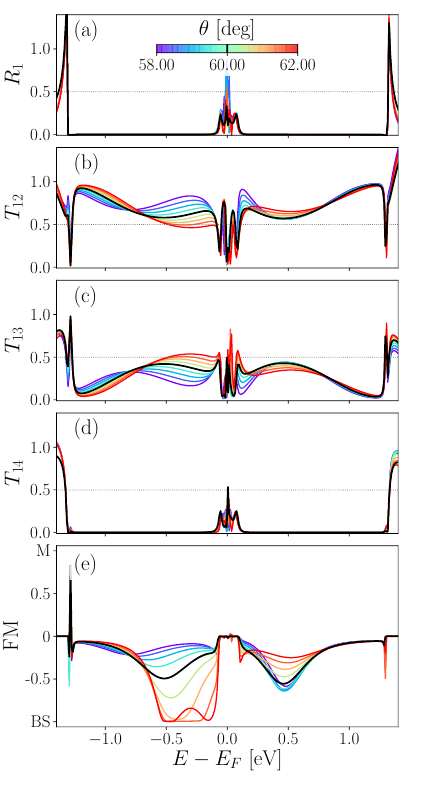}
	\caption{Variation with respect to the rotation angle between two AB-stacked 8-ZGNRs.
	Reflection and transmission probabilities, (a) $R_{1}$ (b) $T_{12}$, (c) $T_{13}$ and (d) $T_{14}$, and (e) figure of merit as a function of the incoming electron energy $E-E_{F}$, obtained for different relative angles ($\theta$) between the ribbons (color lines).
	The reference probabilities ($\theta=60^\circ$) are plotted in black solid lines.}
    \label{fig:var-rotation}
\end{figure}

\begin{figure}
	\includegraphics[width=\columnwidth]{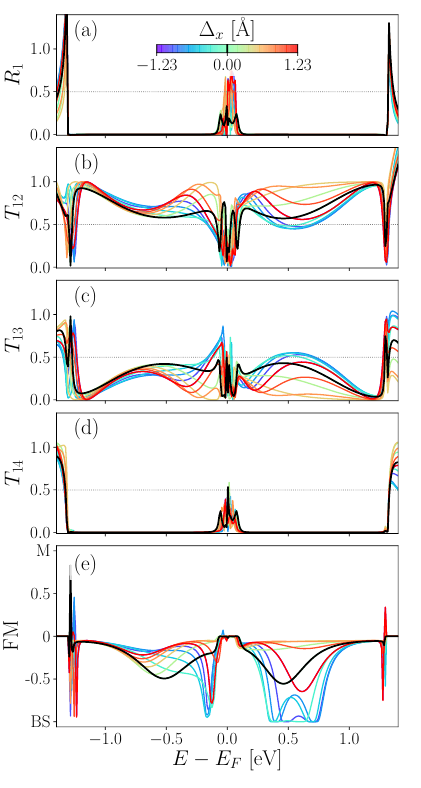}
	\caption{Variation with respect to the relative lateral displacement between two AB-stacked 8-ZGNRs.
	Reflection and transmission probabilities, (a) $R_{1}$ (b) $T_{12}$, (c) $T_{13}$ and (d) $T_{14}$, and (e) figure of merit as a function of electron energy $E-E_{F}$, obtained for different translation distances along the $x$-axis ($\Delta_{x}$) of the on-top ribbon (color lines).
	The reference probabilities ($\Delta_{x}=0$) are plotted in black solid lines.}
    \label{fig:translation-x-axis}
\end{figure}

\begin{figure}
	\includegraphics[width=\columnwidth]{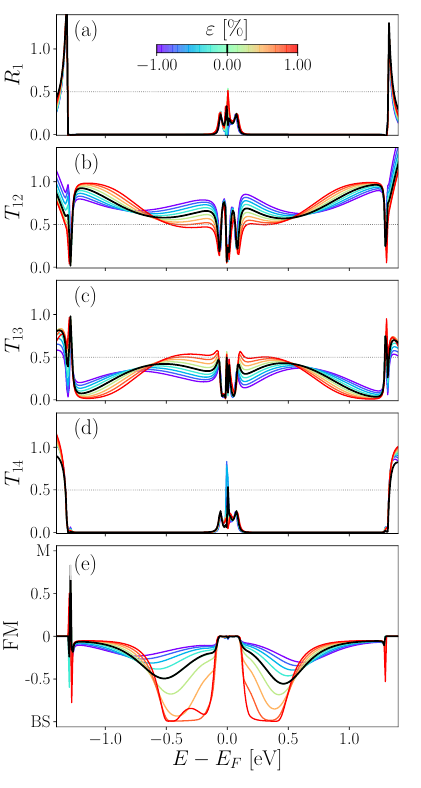}
	\caption{Variation with respect to the applied uniaxial strain $\varepsilon$ along the periodic direction of each GNR for the two AB-stacked 8-ZGNRs.
	Reflection and transmission probabilities, (a) $R_{1}$ (b) $T_{12}$, (c) $T_{13}$ and (d) $T_{14}$, and (e) figure of merit as a function of electron energy $E-E_{F}$, obtained for different uniaxial strain $\varepsilon$ applied to both GNRs along the non-confined direction (color lines).
	The reference probabilities ($\varepsilon=0$) are plotted in black solid lines.}
    \label{fig:strain}
\end{figure}

\begin{figure}
	\includegraphics[width=\columnwidth]{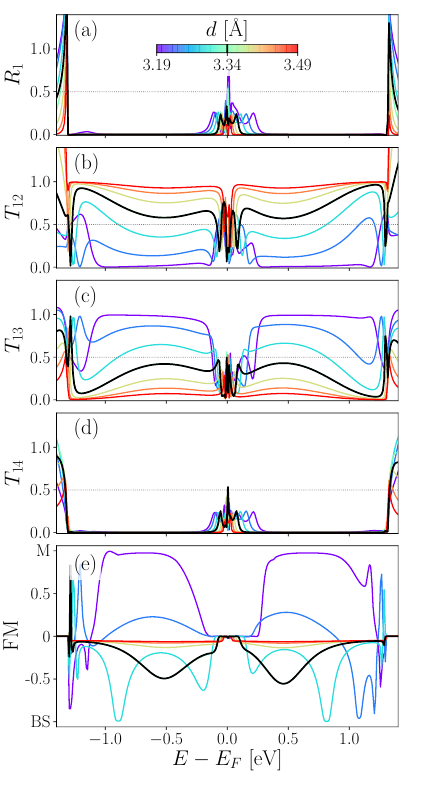}
	\caption{Variation with respect to the \emph{inter}-GNR separation of two AB-stacked 8-ZGNRs.
	Reflection and transmission probabilities, (a) $R_{1}$ (b) $T_{12}$, (c) $T_{13}$ and (d) $T_{14}$, and (e) figure of merit as a function of electron energy $E-E_{F}$ and  \textit{inter}-GNR separation $d$ (color lines).
	The reference probabilities ($d=3.34$ \AA) are plotted in black solid lines.
	}
    \label{fig:inter-distance}
\end{figure}

\begin{figure}
	\includegraphics[width=\columnwidth]{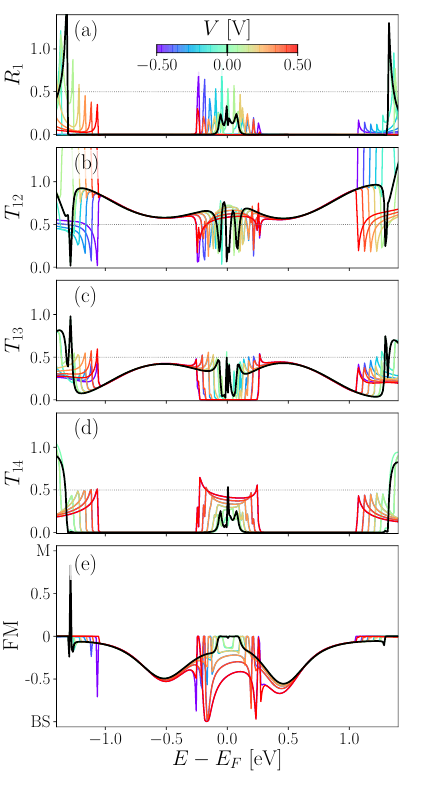}
	\caption{Variation with respect to potential differences $V$ between the two AB-stacked 8-ZGNRs.
	Reflection and transmission probabilities, (a) $R_{1}$ (b) $T_{12}$, (c) $T_{13}$ and (d) $T_{14}$, and (e) figure of merit as a function of electron energy $E-E_{F}$, obtained for different values of $V$ (colored lines).
	The reference probabilities ($V=0$) are plotted in black solid lines.
	}
    \label{fig:var-voltage}
\end{figure}

\subsubsection{Intersection angle}
We first discuss the effect of small rotations of the on-top ribbon starting from the ideal configuration where $\theta=60^{\circ}$.
For instance, the twisting angle between the ribbons introduces separated domains of weakly and strongly coupled atoms in the crossing area that might affect the transport properties of these junctions \cite{Abdullah2017}.
To isolate the effect of the intersection angle from that of the precise stacking pattern (translation), we apply the rotation around the center of the scattering region (crossing) indicated with a black dot in \Figref{fig:geometrical-distortions}(a).
This ensures that the center of the junction remains unchanged and the effect of the rotation angle perturbs mostly the edge zones of the crossing.

Figure \ref{fig:var-rotation} shows the reflection and transmission probabilities for varying angles $\delta\theta = \pm 2^{\circ}$.
The results for the reference case of $\theta=60^{\circ}$ is shown as black lines in all panels.
We first note that the reflection probability $R_1$ does not vary much from its initial value $\sim 0$.
The same holds for the (unwanted) transmission $T_{14}$.
The main effect is the precise distribution between the transmissions $T_{12}$ and $T_{13}$.

This shows that the angle can be a physical knob to tune the transmission ratio between the two outgoing terminals of a BS.
On the other hand, the approximate particle-hole symmetry found for the ideal AB- or AA-stacking goes away as the lattice mismatch grows.
The reflection plane shown in \Figref{fig:devices}a is also lost for $\delta\theta\neq 0$ (and other geometrical distortions), however we still identify only 4 qualitative independent elements in $\mathbf{T}$ for all cases.

\subsubsection{Lateral translations}
	
To study the precise lattice matching in the crossing area, we performed a series of calculations where the top GNR is translated by $\Delta_x$ along the $x$-axis with respect to the bottom GNR, see \Figref{fig:geometrical-distortions}(b).
Due to periodicity it is sufficient to consider translation vectors with modulus $\Delta_{x}\leq 2a\sin(60^{\circ}) \approx 2.46$ {\AA}.

Figure \ref{fig:translation-x-axis} shows the reflection and transmission probabilities as a function of such translations.
Again, the results for the ideal AB-stacking is shown as black lines.
As for small variations in the intersection angle, even though this geometrical distortion also intensifies the particle-hole asymmetry as the system goes away from the ideal stacking, $R_1$ and $T_{14}$ remain rather unaffected by translation.
In other words, the low-loss situation of these devices is robust with respect to translations.
On the other hand, the \textit{inter}-ribbon transfer process of electrons becomes mostly less effective.
We interpret this as due to an overall elongation of interlayer atom distances.
For this reason $T_{13}$ slightly decreases with the translating of the on-top ribbon, while $T_{12}$ slightly increases with respect to the reference curves (black lines) for most of the cases.

\subsubsection{Uniaxial strain}
For experimentally grown GNRs it is relevant to consider the strain-induced deformations, \eg, a lattice mismatch with the supporting substrate \cite{Ni2008}.
But strain can also be applied in a controlled way \cite{Si2016} for example  using a piezoelectric substrate for shrinking or elongating samples by applying a bias voltage \cite{Ding2010}.
In these directions we study here a simplified scenario of applying the same uniaxial strain $\varepsilon$ to both GNRs in the device as defined in \Figref{fig:geometrical-distortions}(c).
As explained in the case of variation of the intersecting angle, to isolate the effect of strain on the transport properties of the device, we apply the strain with respect to the center of the crossing area (as depicted in \Figref{fig:geometrical-distortions}c).
Otherwise arbitrary lattice mismatches could further modify the transmission curves.
The main effect of uniaxial strain is that it induces an anisotropy between the atomic bonds and therefore in the electronic structure of the individual GNRs.
Additionally, a strain induces some mismatch of the lattices in the crossing region, and therefore changes the scattering potential.
The transport properties of the devices are therefore expected to be sensitive to strain.
Figure \ref{fig:strain} explores uniaxial strain in the range from $-1\%$ (compression) to $1\%$ (stretching).
Again, both $R_{1}$ and $T_{14}$ are not affected by the lattice deformation, and remain very close to zero in the single-channel energy region.

Looking at the \textit{intra}- and \textit{inter}-transmissions $T_{12}$ and $T_{13}$ the curves vary smoothly around the reference values (black lines).
The effects of compression and stretching of the GNRs is quite different:
GNR compression causes $T_{12}$ ($T_{13}$) to increase (decrease), while stretching has the opposite effect. 
Again, strain can be seen as a physical knob to engineer the device properties.
For instance, a strain of $\varepsilon\sim 1\%$ brings the system closer to the ideal BS with $T_{12} = T_{13} = 50\%$, while keeping both $R_{1}\sim T_{14}\sim 0$.
In fact, our FM graph of panel \Figref{fig:strain}e shows a significant enhancement of the performance of the device as a BS when stretching the device.

\subsubsection{Interlayer separation}
	
The exponential distance-dependence of electron transport in the tunneling regime suggests that the separation between ribbons may considerably affect the transport properties.
Figure \ref{fig:inter-distance} shows the reflection and transmission probabilities as a function of the GNR separation $d$ within an interval determined by $\pm 0.15$ \AA \ around a typical van der Waals distance $d=3.34$ \AA \ \cite{Brandimarte2017, Baskin1955, Zhao1989} (black lines in all panels).
Apart from the flat-band energy region very close to $E=E_F$, the loss channels characterized by $R_1$ and $T_{14}$ are largely unaffected.

The main effect of varying $d$ is to control the ratio between the \textit{intra}- and \textit{inter}-transmissions $T_{12}$ and $T_{13}$, which varies smoothly to almost 0:1 as the ribbon separation $d$ is decreased.
In the other direction, the ratio goes (unsurprisingly) to 1:0 as the ribbon separation is increased and therefore eventually become decoupled.

The strong variation with the \textit{inter}-GNR separation suggests that this is a key parameter to tune the transport properties.
An ideal 50:50 BS may thus be obtained by applying an external force to the junction for $d\sim 3.30$ \AA.
While a perfect M is found for $d<3.20$ \AA, as seen in \Figref{fig:inter-distance}e, where the plateaus at $\mathrm{FM}=1$ show this behavior.
The possibility to use such electromechanical switching has been also proposed to be used for suspended multilayer graphene \cite{Shi2012}, crossed AGNRs \cite{Brandimarte2017} and crossed carbon nanotubes \cite{Yoon2001}.

\subsubsection{Electrostatic potential differences}
Here we discuss the effect of an electrostatic potential difference between the two ribbons.
This could for instance correspond to an experimental situation where a bias voltage is applied to the GNR electrodes.
We consider a potential difference $V$ that modifies uniformly the onsite energies to $\epsilon_{i}-E_F=-V/2$ (and consequently the chemical potentials of the electrodes) of the top (red) ribbon and $\epsilon_{i}-E_F=V/2$ of the bottom (blue) ribbon (see \Figref{fig:definition-terminals}).

Figure \ref{fig:var-voltage} shows the reflection and transmission probabilities for the range $|V|\leq 0.5$ V.
Drastic changes are observed in the energy range between the electrode chemical potentials, $[-V/2, V/2]$, 
where valence bands (VB) and conduction bands (CB) of the two GNRs now overlap.
In fact, the mixing of VB and CB leads to an interchange of the propagation direction: A transferred electron in the bias window turns 60$^\circ$ instead of 120$^\circ$.
In fact, our FM (\Figref{fig:var-voltage}e) shows that the performance of the device is enhanced in the energy window $|E - E_{F}| \leq V/2$, compared to the unbiased case (black curve).
In contrast, the single-channel energy region slightly shrinks, as the chemical potential shifting produces the edge of the single-mode part of the CB (VB) of the bottom (top) ribbon to overlap with more than one mode in the top (bottom) ribbon.
The presence of multiple bands in any of the incoming or outgoing electrodes is responsible for the larger reflection and transmission into the other output, \eg, as it happens for energies $|E-E_{F}|> 1.0$ eV in panels \Figref{fig:var-voltage}(a,d).

Outside the bias window the curves are hardly changed, reflecting a low variability of the transport properties even when the elastically transferred electron wave to the other ribbon is now propagating with a different momentum due to the energy offsets of their band structures.

\section{Conclusions and outlook}
\label{sec:conclusion}

In this paper we studied the electronic transport properties of 4-terminal devices formed of two intersecting GNRs with a nominal crossing angle of $\theta=60^\circ$.
We presented a complete classification and characterization of the different functionalities that can be found in these type of junctions by varying the edge topology of the GNRs (zigzag or armchair), stacking sequence (AA or AB), width of the ribbons, and energy for the propagating electrons in the valence or conduction bands.

We determined the number of independent transmission probability matrix elements in \Eqref{eq:Tpq-matrix} that fully characterize their transport behavior: 10, 6 or 4 depending on the degree of symmetry that a given device displays.
In practice, however, we found that for low-energy electrons it suffices qualitatively to describe the transmission probabilities with only 4 independent elements.
The reason for this is the fact that the dominant part of the scattering potential contains more symmetries than that of the device geometry as a whole.
Implicitly, this result also means that the strict geometrical symmetry behind the systems is not critical for the GNR crossings to function as beam splitters.

Besides the BS property, we also identified other interesting electron quantum optical functionalities of these devices.
For instance, depending on the GNR width and electron energy the device can also behave as a mirror or an energy filter.

Interestingly, for AA-stacked AGNRs we discovered that there exist two different configurations (\texttt{AA-1} and \texttt{AA-2}) that show little geometrical difference but behave very differently from each other in terms of the electron transport for low-energy electrons.
In the particular case of $3p+2$-AGNR crossings, the electron beam is only allowed to turn $60^\circ$ for the \texttt{AA-1} configuration, as opposed to to $120^\circ$ for the \texttt{AA-2} configuration.
On the other hand, AB-stacked AGNR devices do not show good electron quantum optical features due to the comparatively larger losses and low inter-GNR transmission.
Unfortunately, AA-stacked configurations are probably harder to realize in practice (not the most stable energetically) compared to the AB-stacked one \cite{Mostaani2015}.
Combined with a generally larger variability of the AGNR transport behavior, these facts indicate that ZGNRs are more interesting objects for the considered device applications than AGNRs.

We further tested the robustness of the predicted transport properties 
by studying small variations on the intersecting angle between the ribbons, lattice matching in the crossing area, uniaxial strain, interlayer separation, and finite potential differences for devices composed of 8-ZGNRs.
While the overall qualitative behavior was found to be robust under these modifications, a strong quantitative response can be obtained - indicating the need to control these effects as well as there potential for \emph{tuning} the crossed-GNR devices.
On the other hand, in this work we considered the situation of a spin-degenerate electronic structure. However, ZGNRs have been predicted to develop spin-polarized states localized along the edges of the ribbons close to the Fermi level \cite{Son2006a}.
This suggests that additional spin-dependent effects could emerge in these devices.
The interplay with the physics discussed here could become an interesting topic for future research.

For electron quantum optics applications, the central feature of the considered devices is that they coherently distribute incoming electrons in the intended output ports.
In our model, with a precisely given unitary scattering matrix and without considering environmental degrees of freedom, all the considered devices process the input coherently.
The analysis of the operative decoherence processes in GNR-based devices is an important task for future work.
In particular, a single pure-state electron injected into one arm of a BS device discussed here is mapped to an (mode-)entangled state of the  output ports.
Such entanglement could be verified experimentally, for example by measuring the state's Bell correlations as discussed in \cite{Dasenbrook2016}.
A second basic application of the BS device is the Hanbury Brown--Twiss setup \cite{Henny1999, Oliver1999, Samuelsson2004, Neder2007}, which can be used to study the indistinguishability of electrons prepared in different input ports by the observation of anti-bunching in the output ports of the BS.
A theoretical analysis of these experiments would include the investigation of the influence of environmental degrees of freedom (phonons, electrons in the substrate, or fluctuating perturbations as the ones discussed in \Secref{sec:robustness}), and, in the case of the Hanbury Brown--Twiss setup, also the effect of the interaction between electrons in the BS.
An important prerequisite for all such experiments are methods to inject single electrons in a well-defined mode and to reliably detect them.

Finally, we envision that the functionalities presented here may be interesting as fundamental building blocks in larger electronic networks based on GNRs.
For instance, with four GNRs one could construct the electronic analogue of the Mach--Zehnder interferometer, consisting of two beam splitters and two oriented mirrors at the intersection of pairwise parallel ribbons.
Such a versatile setup, sensitive to the relative phase shift between two spatially separated pathways, have a wide range of quantum technology applications, e.g., metrology, entanglement, cryptography, and information processing \cite{Sarkar2006}.

\begin{acknowledgments}
This work was supported by the project Spanish Ministerio de Econom\'ia y Competitividad (MINECO) through the Grants no.~FIS2017-83780-P (Graphene Nanostructures ``GRANAS'') and no.~MAT2016-78293-C6-4R, the Basque Departamento de Educaci\'on through the PhD fellowship no.~PRE\_2019\_2\_0218 (S.S.),  the University of the Basque Country through the Grant no.~IT1246-19, and the European Union (EU) through Horizon 2020 (FET-Open project SPRING Grant no.~863098).
\end{acknowledgments}

\appendix

\section{Comparison with DFT calculations}\label{sec:supp-DFT-fitting}
	
In this appendix we compare the results presented in the main text with DFT, another popular theoretical approach used in the field of solid state physics.
In particular, we choose to compute the electronic structure of AB-stacked bilayer graphene as a model system to establish suitable parameters for the general TB Hamiltonian introduced in \Secref{sec:methodology}.
We further simulate the electron transport characteristics of the specific device geometries shown in \Figref{fig:devices} for detailed bench-marking. 

We employ the self-consistent DFT and NEGF methods as implemented in the \textsc{Siesta}/\textsc{TranSiesta} \cite{Soler2002, Brandbyge2002, Papior2017} packages.
All calculations of this kind used the vdW density functional \cite{Dion2004} with the modified exchange by Klime\v s \textit{et al.} \cite{Klimes2009}.
The core electrons were described with Troullier-Martins pseudopotentials \cite{Troullier1991} and a double-$\zeta$ basis set defined with a 30 meV energy shift was used to expand the valence-electron wave functions.
The fineness of the real-space integration mesh was defined using a 350 Ry energy cutoff.
All carbon atoms were saturated at the edges with hydrogen atoms.

\begin{figure}[h!]
	\includegraphics[width=\columnwidth]{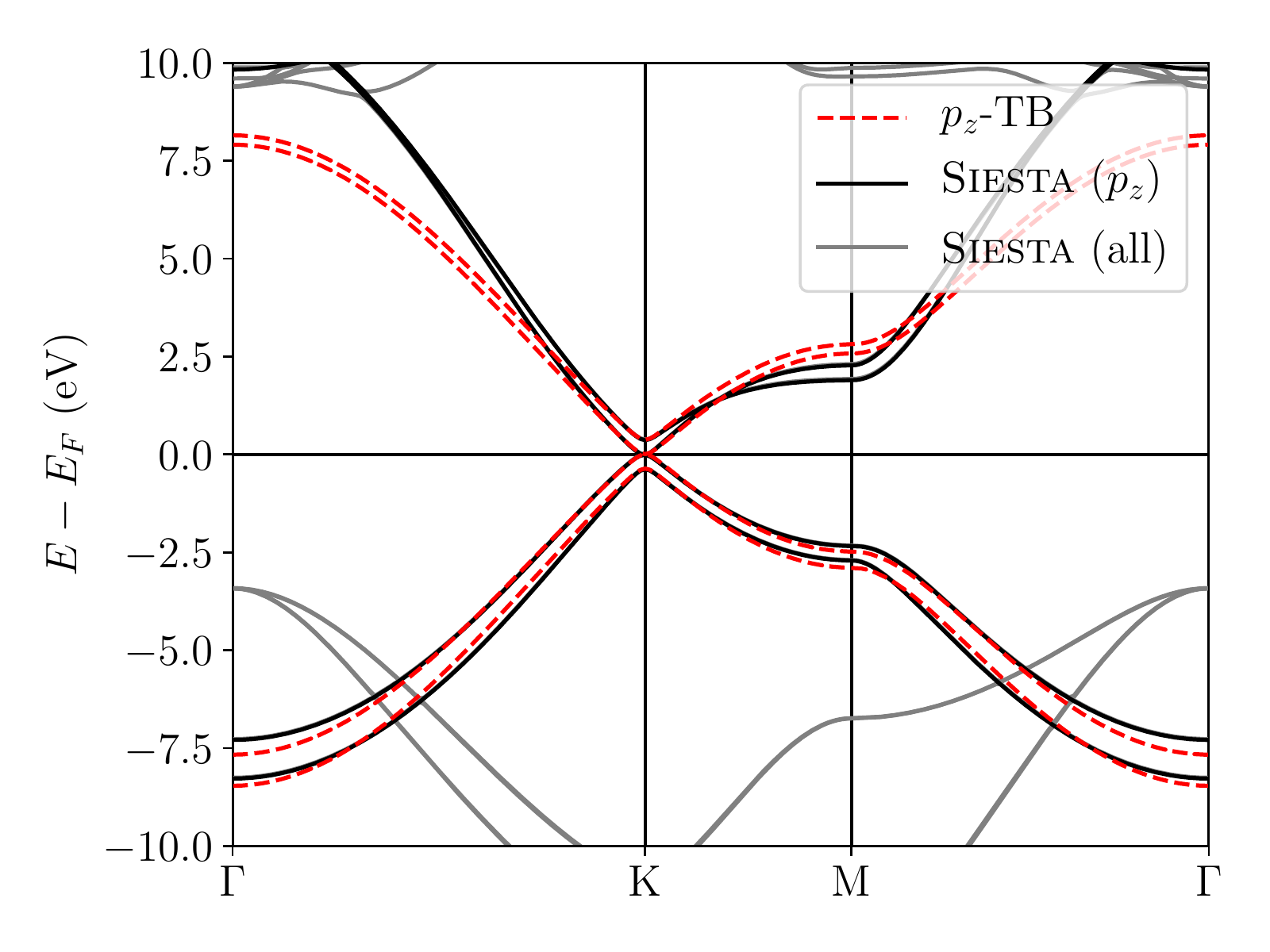}
	\caption{Band structure of AB-stacked bilayer graphene along the $\Gamma$--K--M--$\Gamma$ path of the Brillouin zone, obtained with DFT (black and gray solid lines) and TB methods (red dashed lines), with the fitted hopping parameters described in the text.
	The bond length is set to $a=1.42$ {\AA} and the interlayer separation to $d=3.34$ \AA.
	Black lines correspond to the graphene $\pi$ bands (formed by the $p_{z}$ orbitals) while the gray lines show the graphene $\sigma$ bands absent in the TB model.}
	\label{fig:bands-AB-bilayer}
\end{figure}

\begin{figure}[t]
    \includegraphics[width=\columnwidth]{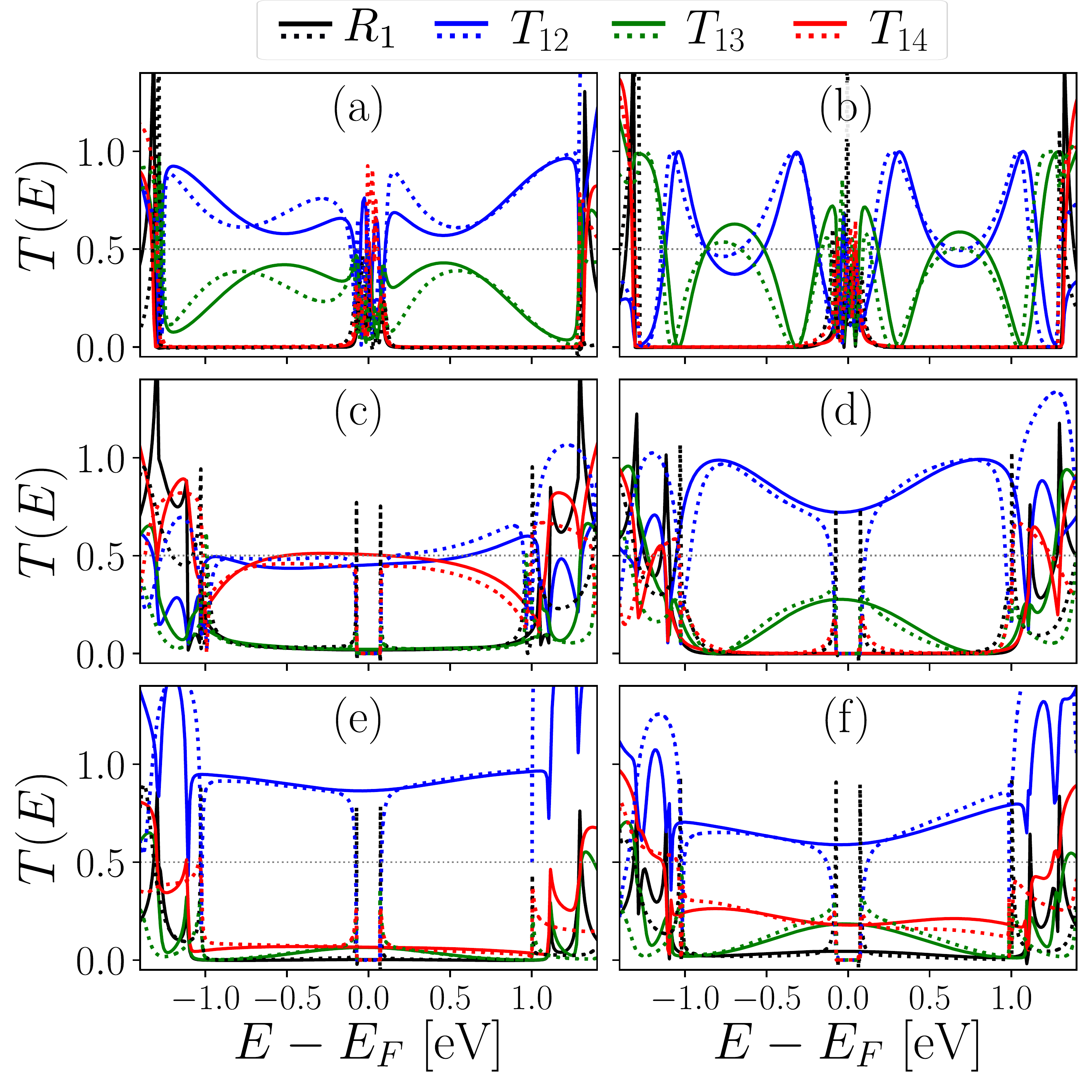}
    \caption{Reflection and transmission probabilities $R_{1}$ (black), $T_{12}$ (blue), $T_{13}$ (green) and $T_{14}$ (red), obtained with both TB (solid lines) and DFT (dotted lines) methods, through the devices of \Figref{fig:devices}:
    two crossed 8-ZGNRs in configuration (a) \texttt{AB} and (b) \texttt{AA}, and two crossed 11-AGNRs in configuration (c) \texttt{AA-1}, (d) \texttt{AA-2}, (e) \texttt{AB-1} and (f) \texttt{AB-2}.
    }
    \label{fig:supp-DFT-vs-TB}
\end{figure}

Figure \ref{fig:bands-AB-bilayer} shows the calculated electronic bands along the $\Gamma$--K--M--$\Gamma$ path of the Brillouin zone of AB-stacked bilayer graphene obtained with \textsc{Siesta} \cite{Soler2002}.
Given the usage of a double-$\zeta$ basis set, the orthogonal $\sigma$ and $\pi$ bands have simple representations in terms of the $\{s, p_x, p_y\}$ and $\{p_z\}$ basis orbitals, respectively.
To map the DFT electronic structure onto the effective TB model in Eqs.~(\ref{eq:TB-Hamiltonian})-(\ref{eq:Vpsig}), it is thus sufficient to consider only the $p_z$ part of the DFT Hamiltonian.
Since we are interested in the low-energy physics, we fitted the TB bands inside an energy window of $|E-E_{F}|\leq 2$ eV using non-linear least squares and obtained the following optimal hopping parameters used in the main text: $t_{\parallel} = 2.682$ eV, $t^{\prime}=2.7$ meV and $t_{\perp} = 0.371$ eV.
The corresponding TB bands with these parameters are plotted in red dashed lines in \Figref{fig:bands-AB-bilayer}, showing a very good agreement in the energy range of relevance in this work.
Albeit unnecessary for the purposes here, we note that the significant deviations at the $\pi$ band edges can readily be improved with a non-orthogonal TB model by introduction of additional parameters for the overlap matrix.

Having fixed the parameters for the TB model we proceed to compare it against the derived transport properties from DFT-NEGF for the six characteristic devices shown \Figref{fig:devices}.
Figure \ref{fig:supp-DFT-vs-TB} shows the computed reflection and transmission probabilities from TB (solid lines) and DFT (dotted lines) within an energy window of $|E-E_F| \lesssim 1.5$ eV.
Apart from different magnitudes of the AGNR band gap (known to be related to edge effects ignored in this TB modeling \cite{Son2006b}) the two models only show minor numerical differences.
Overall the two models provide very similar shapes and quantitative results for the  transmission functions.
From \Figref{fig:bands-AB-bilayer} and \Figref{fig:supp-DFT-vs-TB} we therefore conclude that the TB method used in the main text provides an accurate description of the essential physics in the energy range we are interested here.

\bibliography{main.bbl}

\includepdf[pages={{},-}, pagecommand={\clearpage \thispagestyle{empty}}, scale=1]{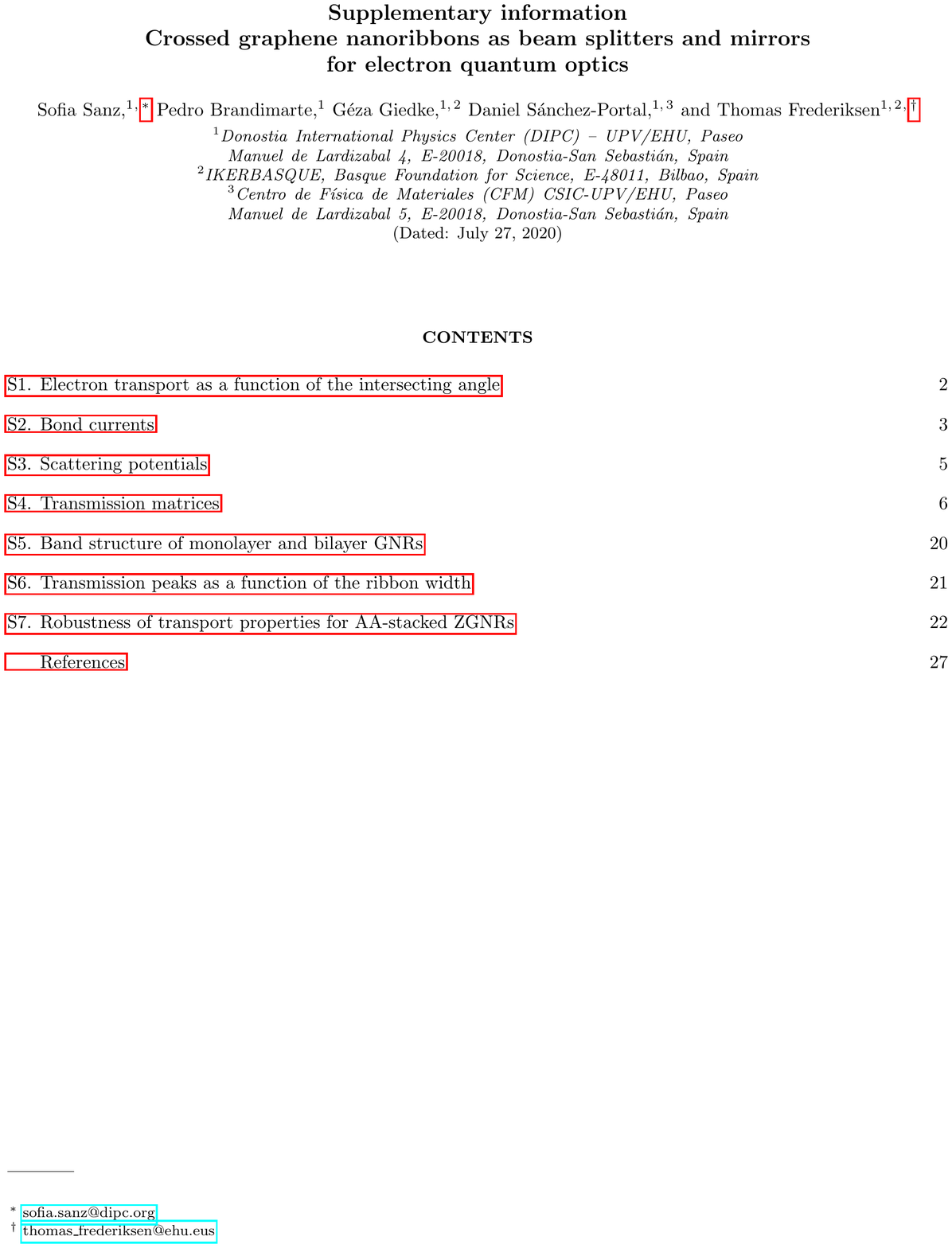}
\thispagestyle{empty}

\end{document}